\documentstyle[11pt]{article}
\title{The Quantum Information Manifold for $\varepsilon$-Bounded Forms}
\author{M. R. Grasselli\thanks{Supported by a grant from CAPES-Brazil.}
\qquad\qquad R. F. Streater \\Dept. of Mathematics,\\King's College,
\\Strand,\\London, WC2R 2LS}
\date{29/09/1999}
\newtheorem{corollary}[equation]{Corollary}
\newtheorem{theorem}[equation]{Theorem}
\newtheorem{lemma}[equation]{Lemma}
\newtheorem{definition}[equation]{Definition}

\begin{document}
\maketitle
\begin{abstract}
Let $H_0 \geq I$ be a self-adjoint operator and let $V$ be a
form-small perturbation such that
$\left\|R_0^{\frac{1}{2}+\varepsilon}VR_0^{\frac{1}{2}-\varepsilon}
\right\|<\infty$, where $\varepsilon \in (0,1/2)$ and $R_0=H_0^{-1}$.
Suppose that there exists a positive $\beta <1$ such that \mbox{$Z:=
\mbox{Tr}e^{-\beta H_0}<\infty$}. Then we show that the free energy
$\Psi=\log Z$ is an analytic fucntion in the sense of Fr\'{e}chet,
and that the family of density operators defined in this way is an
analytic manifold.
\end{abstract}

The use of differential geometric methods in parametric estimation
theory is by now a fairly sound subject, whose foundations,
applications and techniques can be found in several books \cite{Amari,
Kass,Murray}. The non-parametric version of this {\em Information
Geometry} had its mathematical basis laid down in recent years
\cite{GibilescoPistone,PistoneSempi}. It is a genuine branch of
infinite-dimensional analysis and geometry. The theory of quantum
information manifolds aims to be its noncommutative counterpart
\cite{Hasagawa2,Nagaoka,PetzSudar,Petz}.

In this paper we generalise the results obtained by one of us
\cite{Streater1,Streater2} to a larger class of potentials. In \S1 we
introduce $\varepsilon$-bounded perturbations of a given Hamiltonian
and review their relation with form-bounded and operator-bounded
perturbations. In \S2 we construct a Banach manifold of quantum
mechanical states with $(+1)$-affine structure and
$(+1)$-connection, using the $\varepsilon$-bounded perturbations.
Finally, in \S3 we prove analyticity of the free energy $\Psi_X$
in sufficiently small neighbourhoods in this manifold, from which
it follows that the $(-1)$-coordinates are analytic.

\section{$\varepsilon$-Bounded Perturbations}

We recall the concepts of operator-bounded and form-bounded
perturbations \cite{Kato}. Given operators $H$ and $X$ defined on dense
domains ${\cal D}(H)$ and ${\cal D}(X)$ in a Hilbert space $\cal H$,
we say that $X$ is {\em $H$-bounded\/} if
\begin{enumerate}
  \item ${\cal D}(H)\subset{\cal D}(X)$ and
  \item there exist positive constants $a$ and $b$ such that
  \[\left\|X\psi\right\|\leq a\left\|H\psi\right\| +
  b\left\|\psi\right\|,\mbox{ for all $\psi\in{\cal
  D}(H)$.}\]
\end{enumerate}
Analogously, given a positive self-adjoint operator $H$ with
associated form $q_H$ and form domain $Q(H)$, we say that a
symmetric quadratic form $X$ (or the symmetric sesquiform
obtained from it by polarization) is {\em $q_H$-bounded\/} if
\begin{enumerate}
  \item $Q(H)\subset Q(X)$ and
  \item there exist positive constants $a$ and $b$ such that
  \[\left|X(\psi,\psi)\right|\leq aq_H(\psi,\psi) +
  b(\psi,\psi),\mbox{ for all $\psi\in Q(H)$.}\]
\end{enumerate}

In both cases, the infimum of such $a$ is called the relative
bound of $X$ (with respect to $H$ or with respect to $q_H$,
accordingly).

Suppose that $X$ is a quadratic form with domain $Q(X)$ and $A,B$
are operators on $\cal H$ such that $A^*$ and $B$ are densely
defined. Suppose further that $A^*:{\cal D}(A^*)\rightarrow Q(X)$
and $B:{\cal D}(B)\rightarrow Q(X)$. Then the expression $AXB$
means the form defined by \[\phi,\psi\mapsto
X(A^*\phi,B\psi),\qquad \phi\in {\cal D}(A^*),\quad \psi\in {\cal
D}(B).\]

With this definition in mind, let us specialise to the case where
$H_0\geq I$ is a self-adjoint operator with domain ${\cal D}(H_0)$,
quadratic form $q_0$ and form domain $Q_0={\cal D}(H_0^{1/2})$, and
let $R_0=H_0^{-1}$ be its resolvent at the origin. Then it is easy
to show that a symmetric operator $X:{\cal D}(H_0)\rightarrow \cal H$ is
$H_0$-bounded if and only if $\left\|XR_0\right\|<\infty$. The
following lemma is also known \cite[lemma 2]{Streater1}.

\begin{lemma}
A symmetric quadratic form $X$ defined on $Q_0$ is $q_0$-bounded if
and only if $R_0^{1/2}XR_0^{1/2}$
is a bounded symmetric form defined everywhere. Moreover, if 
$\left\|R_0^{1/2}XR_0^{1/2}\right\|<\infty$
then the relative bound $a$ of $X$ with respect to $q_0$ satisfies
$a\leq \left\|R_0^{1/2}XR_0^{1/2}\right\|$.
\end{lemma}

The set ${\cal T}_{\omega}(0)$ of all $H_0$-bounded symmetric
operators X is a Banach space with norm $\|X\|_\omega
(0) :=\left\|XR_0\right\|$, since the map $A\mapsto AH_0$ from $\cal B(H)$
onto ${\cal T}_{\omega}(0)$ is an isometry.

The set ${\cal T}_0(0)$ of all $q_0$-bounded symmetric forms $X$
is also a Banach space with norm
$\|X \|_0 (0):=\left\|R_0^{1/2}XR_0^{1/2}\right\|$, since
the map $A\mapsto H_0^{1/2}AH_0^{1/2}$ from the set of all bounded
self-adjoint operators on $\cal H$ onto ${\cal T}_0(0)$ is again
an isometry.

Now, for $\varepsilon\in (0,1/2)$, let ${\cal T}_{\varepsilon}(0)$
be the set of all symmetric forms X with ${\cal D}(H_0^{
\frac{1}{2}-\varepsilon}) \subset Q(X)$ and such that $\|X \|_\varepsilon
(0):=\left\|R_0^{\frac{1}{2}+\varepsilon}XR_0^{\frac{1}{2}-\varepsilon}
\right\|$
is finite. Then the map $A\mapsto H_0^{\frac{1}{2}-\varepsilon}AH_0^
{\frac{1}{2}+\varepsilon}$
is an isometry from the set of all bounded self-adjoint operators
on $\cal H$ onto ${\cal T}_{\varepsilon}(0)$. Hence ${\cal T}_{\varepsilon}
(0)$ is a Banach space with the $\varepsilon$-norm
$\|\cdot\|_\varepsilon (0)$. We note that $ {\cal D}(H_0^{\frac{1}{2}})
\subset{\cal D}(H_0^{\frac{1}{2}-\delta})$, for all $0 \leq \delta \leq
1/2$.

We can now prove the following lemma.

\begin{lemma}
For fixed symmetric $X$, $\left\|X\right\|_\varepsilon$ is a
monotonically increasing function of $\varepsilon\in [0,1/2]$.
\label{monotonicity}
\end{lemma}
{\em Proof:} We have to prove that $\left\|R_0^yXR_0^{1-y}\right\|$ is
increasing for $y\in [1/2,1]$ and decreasing for $y\in [0,1/2]$.
Let $\frac{1}{2} \leq \delta \leq 1$ and suppose that
\mbox{$\left\|R_0^{\delta}XR_0^{1-\delta}\right\|<\infty$}. Interpolation
theory for Banach spaces \cite{ReedSimon} and the fact that $\left\|R_0^
{\delta}XR_0^{1-\delta}\right\|=\left\|R_0^{1-\delta}XR_0^{\delta}\right\|$
then give \[\left\|R_0^xXR_0^{1-x}\right\|\leq
\left\|R_0^{\delta}XR_0^{1-\delta}\right\|,\mbox{      for all }x\in
[1-\delta,\delta],\] and particularly for $\frac{1}{2}\leq y\leq \delta
\leq 1$, we have \[\left\|R_0^yXR_0^{1-y}\right\|\leq
\left\|R_0^{\delta}XR_0^{1-\delta}\right\|.\]
By the other hand, for $0 \leq 1-\delta \leq y\leq \frac{1}{2}$,
\[\left\|R_0^yXR_0^{1-y}\right\|\leq
\left\|R_0^{\delta}XR_0^{1-\delta}\right\|=\left\|R_0^{1-\delta}X
R_0^{\delta}\right\|. \quad \Box \]

\section{Construction of the Manifold}

\subsection{The First Chart}

Let ${\cal C}_p,0<p<1$, denote the set of compact operators $A:\cal H
\mapsto\cal H$ such that $|A|^p \in {\cal C}_1$, where ${\cal C}_1$ is the
set of trace-class operators on $\cal H$. Define \[{\cal
C}_{<1}:=\bigcup_{0<p<1} {\cal C}_p. \]

We take the underlying set of the quantum information manifold to
be \[{\cal M=C}_{<1}\cap\Sigma\] where $\Sigma \subseteq {\cal C}_1$
denotes the set of density operators. We do so because the next
step of our project is to look at the Orlicz space geometry
associated with the quantum information manifold
\cite{GibilescoPistone} and the quantum analogue of classical Orlicz
space $L\log L$ seems to be \[{\cal C}_1 \! \log{\cal C}_1:=\{\rho \in
{\cal C}_1 : S(\rho)=-\sum \lambda_i \log \lambda_i <\infty\},\]
where $\{\lambda_i\}$ are the singular numbers of $\rho$. With
this notation, the set of normal states of finite entropy is
${\cal C}_1 \! \log {\cal C}_1 \cap \Sigma$
and we have ${\cal C}_{<1} \subset {\cal C}_1 \! \log {\cal C}_1$. At
this level, $\cal M$ has a natural affine structure defined as
follows: let $\rho_1 \in {\cal C}_{p_1}\cap \Sigma$ and $\rho_2 \in
{\cal C}_{p_2}\cap\Sigma$; take $p=\max \{p_1,p_2\}$, then $\rho_1,\rho_2
\in {\cal C}_{p}\cap\Sigma$, since $p\leq q$ implies ${\cal C}_p \subseteq
{\cal C}_q$ \cite{Pietsch}; define ``$\lambda\rho_1+(1-\lambda)\rho_2, 0
\leq\lambda \leq 1$'' as the usual sum of operators in ${\cal C}_p$. This
is called the $(-1)$-affine structure.

We want to cover ${\cal M}$ by a Banach manifold. In \cite{Streater1}
this is achieved defining hoods of $\rho \in {\cal M}$ using
form-bounded perturbations. The manifold obtained there is shown to
have a Lipschitz structure. In \cite{Streater2} the same is done
with the more restrictive class
of operator-bounded perturbations. The result then is that the
manifold has an analytic structure. We now proceed using
$\varepsilon$-bounded perturbations, with a similar result.

To each $\rho_0 \in {\cal C}_{\beta_0}\cap \Sigma$, $\beta_0 <
1$, let $H_0=-\log \rho_0 + cI \geq I$ be a self-adjoint operator
with domain ${\cal D}(H_0)$ such  that \begin{equation}
\rho_0=Z_0^{-1}e^{-H_0}=e^{-(H_0+\Psi_0)}.
\end{equation}

In ${\cal T}_\varepsilon (0)$, take $X$ such that 
$\|X\|_\varepsilon (0) < 1-\beta_0$. Since $\|X \|_0 (0) \leq \|X
\|_\varepsilon (0) <1-\beta_0$, $X$ is also $q_0$-bounded with bound $a_0$
less than $1-\beta_0$. The {\em KLMN} theorem then tells us that there
exists a unique semi-bounded self-adjoint operator $H_X$ with form $q_X =
q_0 + X$ 
and form domain $Q_X = Q_0$. Following an unavoidable abuse of
notation, we write $H_X=H_0+X$ and consider the operator \begin{equation}
\rho_X = Z_X^{-1}e^{-(H_0+X)}=Z_X^{-1}e^{-(H_0+X+\Psi_X)}.
\end{equation}

Then $\rho_X \in {\cal C}_{\beta_X}\cap \Sigma$, where $\beta_X =
\frac{\beta_0}{1-a_0}<1$
\cite[lemma 4]{Streater1}. The state $\rho_X$ does not change if we
add to $H_X$ a multiple of the identity in such a way that $H_X+cI
\geq I$, so we can always assume that, for the perturbed state, we have
$H_X \geq I$. We take as a hood ${\cal M}_0$ of $\rho_0$ the set of all
such states, that is, ${\cal M}_0=\{\rho_X:\|X \|_\varepsilon (0)
<1-\beta_0\}$.

Because $\rho_X = \rho_{X+\alpha I}$, we introduce in ${\cal T}_\varepsilon
(0)$ the equivalence relation \mbox{$X \sim Y$} iff $X-Y=\alpha I$ for some
$\alpha \in{\bf R}$. We then identify $\rho_X$ in ${\cal M}_0$ with the
line $\{Y \in {\cal T}_\varepsilon (0): Y=X+\alpha I, \alpha \in {\bf R}\}$
in ${\cal T}_\varepsilon (0)/\! \sim$. This is a bijection from ${\cal M}_0$
onto the subset of ${\cal T}_\varepsilon (0)/\! \sim$ defined by
$\left\{ \{X+\alpha I\}_{\alpha \in {\bf R}}:\|X \|_\varepsilon (0)< 1-\beta_0
\right\}$ and ${\cal M}_0$ becomes topologised by transfer of structure. Now
that ${\cal M}_0$ is a (Hausdorff) topological space, we want to
parametrise it by an open set in a Banach space. By analogy with
the finite dimensional case \cite{PetzToth,Hasagawa1,Nagaoka}, we want to use
the Banach subspace of centred variables in ${\cal T}_\varepsilon (0)$; in
our terms, perturbations with zero mean (the `scores'). For this, define
the regularised mean of $X \in {\cal T}_\varepsilon (0)$ in the state
$\rho_0$ as \begin{equation}
\rho_0\! \cdot \! X :=Tr(\rho_0^\lambda X
\rho_0^{1-\lambda}),\qquad \quad \mbox{for $0<\lambda <1$}.
\end{equation}

Since $\rho_0 \in {\cal C}_{\beta_0}\cap \Sigma$ and $X$ is
$q_0$-bounded, lemma 5 of \cite{Streater1} ensures that $\rho_0 \!
\cdot \! X$ is finite and independent of $\lambda$. It was a shown there
that $\rho_0 \! \cdot \! X$ is a continuous map from ${\cal T}_0 (0)$
to ${\bf R}$, because its bound contained a factor
$\|X \|_0 (0)$. Exactly the same proof shows that $\rho_0 \! \cdot \! X$
is a continuous map from ${\cal T}_\varepsilon (0)$ to ${\bf
R}$. Thus the set $\widehat{\cal T}_\varepsilon (0):=
\left\{X \in {\cal T}_\varepsilon (0):\rho_0 \! \cdot \! X=0 \right\}$
is a closed subspace of ${\cal T}_\varepsilon (0)$ and so is a
Banach space with the norm $\left\| \cdot \right\|_\varepsilon$
restricted to it.

To each $\rho_X \in {\cal M}_0$, consider the unique intersection of the
equivalence class of $X$ in ${\cal T}_\varepsilon (0)/\! \sim$ with the set
$\widehat{\cal T}_\varepsilon (0)$, that is, the point in the line
$\{X+\alpha I \}_{\alpha \in {\bf R}}$  with $\alpha=-\rho_0 \! \cdot \! X$.
Write $\widehat{X}=X-\rho_0 \! \cdot \! X$ for this point. The map $\rho_X
\mapsto\widehat{X}$
is a homeomorphism between ${\cal M}_0$ and the open subset of
$\widehat{\cal T}_\varepsilon (0)$ defined by $\left\{\widehat{X}:
\widehat{X}=X-\rho_0 \! \cdot \! X, \left\|X\right\|_\varepsilon <1-\beta_0
\right\}$. The map $\rho_X \mapsto \widehat{X}$ is then a global chart
for the Banach manifold ${\cal M}_0$ modeled by $\widehat{\cal T}_\varepsilon
(0)$. As usual, we identify the tangent space at $\rho_0$ with
$\widehat{\cal T}_\varepsilon
(0)$, the tangent curve $\left\{\rho (\lambda)=Z_{\lambda X}^{-1}e^{-(H_0+
\lambda X)}, \lambda \in [-\delta ,\delta]\right\}$
being identified with $\widehat{X}=X-\rho_0 \! \cdot \! X$.

\subsection{Enlarging the Manifold}

We extend our manifold by adding new patches compatible with
${\cal M}_0$. The idea is to construct a chart around each perturbed
state $\rho_X$ as we did around $\rho_0$. Let $\rho_X \in {\cal M}_0$
with Hamiltonian $H_X \geq I$ and consider the Banach space ${\cal T}
_\varepsilon (X)$ of all symmetric forms $Y$ such that the norm 
$\|Y\|_\varepsilon (X):= \left\|R_X^{\frac{1}{2} + \varepsilon}
YR_X^{\frac{1}{2} - \varepsilon}\right\|$
is finite, where $R_X=H_X^{-1}$ denotes the resolvent of $H_X$ at the origin.
In ${\cal T}_\varepsilon (X)$, take $Y$ such that $\|Y\|_\varepsilon (X)<1
-\beta_X$. From lemma~\ref{monotonicity} we know that Y is $q_X$-bounded
with bound $a_X$ less than $1 - \beta_X$. Let $H_{X+Y}$ be the unique
semi-bounded self-adjoint operator, given by the {\em KLMN} theorem, with
form $q_{X+Y} = q_X + Y = q_0 + X + Y$ and form domain $Q_{X+Y}=Q_X=Q_0$.
Then the operator
\begin{equation}
\rho_{X+Y}=Z_{X+Y}^{-1}e^{-H_{X+Y}}=Z_{X+Y}^{-1}e^{-(H_0+X+Y)}
\end{equation}
is in ${\cal C}_{\beta_Y}\cap \Sigma$, where $\beta_Y=
\frac{\beta_X}{1-a_X}$.

We take as a neighbourhood of $\rho_X$ the set ${\cal M}_X$ of
all such states. Again $\rho_{X+Y} = \rho_{X+Y+ \alpha I}$, so we furnish
${\cal T}_\varepsilon (X)$ with
the equivalence relation $Z \sim Y$ iff $Z-Y=\alpha I$ and we see
that ${\cal T}_\varepsilon (X)$ is mapped bijectively onto the set
of lines $\left\{\{Z=Y+\alpha I\}_{\alpha \in {\bf R}},\|Y\|_\varepsilon (X)
< 1-\beta_X \right\}$ in ${\cal T}_\varepsilon (X)/ \! \sim$. In this way
we topologise ${\cal
M}_X$, by transfer of structure, with the quotient topology of ${\cal T}
_\varepsilon (X)/ \! \sim$.

Again we can define the mean of $Y$ in the state $\rho_X$ by
\begin{equation}
\rho_X\! \cdot \! Y :=Tr(\rho_X^\lambda Y
\rho_X^{1-\lambda}),\qquad \quad \mbox{for $0<\lambda <1$}.
\end{equation} and notice that it is finite and independent of
$\lambda$. This is a continuous function of $Y$ with respect to
the norm $\| \cdot \|_\varepsilon (X)$, hence $\widehat{\cal T}_
\varepsilon (X)=\{Y \in {\cal T}_\varepsilon (X):\rho_X \! \cdot \! Y=0 \}$
is closed and so is a Banach space with the norm $\|\cdot\|_\varepsilon(X)$
restricted to it. Finally, let $\widehat Y$ be the unique
intersection of the line $\{Z=Y+\alpha I \}_{\alpha \in {\bf R}}$
with the hyperplane $\widehat{\cal T}_\varepsilon (X)$, given by $\alpha
= -\rho_X \! \cdot \! Y$. Then $\rho_{X+Y} \mapsto \widehat Y$ is a
homeomorphism between ${\cal M}_X$ and the open subset of $\widehat
{\cal T}_\varepsilon (X)$ defined by $\left\{\widehat Y \in \widehat{\cal
T}_\varepsilon (X): \widehat Y=Y-\rho_X \! \cdot \! Y, \|Y\|_\varepsilon
(X)<1-\beta_X\right\}$. We obtain that $\rho_{X+Y} \mapsto \widehat Y$ is
a global chart for the manifold ${\cal M}_X$ modeled by $\widehat{\cal T}_
\varepsilon(X)$. The tangent space at $\rho_X$ is identified with $\widehat
{\cal T}_\varepsilon(X)$ itself.

We now look to the union of ${\cal M}_0$ and ${\cal M}_X$. We need
to show that our two previous charts are compatible in the
overlapping region ${\cal M}_0 \cap {\cal M}_X$. But first we prove the
following series of technical lemmas.

\begin{lemma}
Let $X$ be a symmetric form defined on $Q_0$ such that \mbox{$\left\|
R_0^{1/2}XR_0^{1/2}\right\| < 1$}. Then ${\cal D} (H_0^{\frac{1}{2}-
\varepsilon})={\cal D}(H_X^{\frac{1}{2}-\varepsilon})$, for any
$\varepsilon\in (0,1/2)$.
\label{domains}
\end{lemma}
{\em Proof:} We know that ${\cal D} (H_0^{1/2})={\cal D}
(H_X^{1/2})$, since $X$ is $q_0$-small. Moreover, $H_X$ and $H_0$
are comparable as forms, that is, there exists $c>0$ such that
\[c^{-1}q_0 (\psi) \leq q_X (\psi) \leq cq_0 (\psi), \qquad \quad
\mbox{for all $\psi \in Q_0$}.\]

Using the fact that $x \mapsto x^\alpha \quad (0\!<\!\alpha\!<\!1)$
is an operator monotone function \cite[lemma 4.20]{Davies}, we conclude that
\[c^{-(1-2\varepsilon)}H_0^{1-2\varepsilon} \leq
H_X^{1-2\varepsilon} \leq c^{1-2\varepsilon}H_0^{1-2\varepsilon},\]
which implies that ${\cal D} (H_0^{\frac{1}{2}-\varepsilon})={\cal
D}(H_X^{\frac{1}{2}-\varepsilon})$.  $\Box$

The conclusion remains true if we now replace $H_X$ by $H_X+I$, if
necessary in order to have $H_X \geq I$. This is assumed in the
next corollary.

\begin{corollary}
The operator $H_X^{\frac{1}{2}-\varepsilon}R_0^{\frac{1}{2}-\varepsilon}$
is bounded and has bounded inverse
$H_0^{\frac{1}{2}-\varepsilon}R_X^{\frac{1}{2}-\varepsilon}$.
\end{corollary}
{\em Proof:} $R_0^{\frac{1}{2}-\varepsilon}$ is bounded and maps $\cal H$
into ${\cal D} (H_0^{\frac{1}{2}-\varepsilon})={\cal D}
(H_X^{\frac{1}{2}-\varepsilon})$. Then $H_X^{\frac{1}{2}-\varepsilon}R_0^
{\frac{1}{2}-\varepsilon}$
is bounded, since $H_X^{\frac{1}{2}-\varepsilon}$ is closed.
By exactly the same argument, we obtain that $H_0^{\frac{1}{2}-\varepsilon}
R_X^{\frac{1}{2}-\varepsilon}$
is bounded. Finally $(H_0^{\frac{1}{2}-\varepsilon}R_X^{\frac{1}{2}-
\varepsilon})(H_X^{\frac{1}{2}-\varepsilon}R_0^{\frac{1}{2}-\varepsilon})
=(H_X^{\frac{1}{2}-\varepsilon}R_0^{\frac{1}{2}-\varepsilon})(H_0^{\frac
{1}{2}-\varepsilon}R_X^{\frac{1}{2}-\varepsilon})=I. \quad \Box$

\begin{lemma}
For $\varepsilon \in (0,1/2)$, let $X$ be a form defined on ${\cal D}
(H_0^{\frac{1}{2}-\varepsilon})$
such that $\left\|R_0^{\frac{1}{2}+\varepsilon}XR_0^{\frac{1}{2}-
\varepsilon}\right\| <1$. Then $R_0^{\frac{1}{2}+\varepsilon}H_X^
{\frac{1}{2}+\varepsilon}$ is bounded and has bounded inverse
$R_X^{\frac{1}{2}+\varepsilon}H_0^{\frac{1}{2}+\varepsilon}$.
Moreover, ${\cal D} (H_0^{\frac{1}{2}+\varepsilon})={\cal D}
(H_X^{\frac{1}{2}+\varepsilon})$
\end{lemma}
{\em Proof:} From lemma~\ref{monotonicity}, we know that $\left
\|R_0^{1/2}XR_0^{1/2}\right\|<1$, so lemma~\ref{domains} and
its corollary apply. We have that
\begin{eqnarray*}
1 & > & \left\|R_0^{\frac{1}{2}+\varepsilon}XR_0^{\frac{1}{2}-\varepsilon}
\right\| \\
&=&\left\|R_0^{\frac{1}{2}+\varepsilon}(H_X-H_0)R_0^{\frac{1}{2}-\varepsilon}
\right\| \\
&=&\left\|R_0^{\frac{1}{2}+\varepsilon}H_XR_0^{\frac{1}{2}-\varepsilon}-
I\right\|,
\end{eqnarray*}
thus $\left\|R_0^{\frac{1}{2}+\varepsilon}H_XR_0^{\frac{1}{2}-\varepsilon}
\right\|<\infty$. We write this as \[\left\|R_0^{\frac{1}{2}+\varepsilon}
H_X^{\frac{1}{2}+\varepsilon}H_X^{\frac{1}{2}-\varepsilon}R_0^
{\frac{1}{2}-\varepsilon}\right\|<\infty.\]

Since $H_X^{\frac{1}{2}-\varepsilon}R_0^{\frac{1}{2}-\varepsilon}$
is bounded and invertible, so is
$R_0^{\frac{1}{2}+\varepsilon}H_X^{\frac{1}{2}+\varepsilon}$.
Finally, the fact that $\left\|R_0^{\frac{1}{2}+\varepsilon}H_X^{\frac{1}
{2}+\varepsilon}\right\|<\infty$ and $\left\|R_X^{\frac{1}{2}+\varepsilon}
H_0^{\frac{1}{2}+\varepsilon}\right\|<\infty$
implies that $H_X^{\frac{1}{2}+\varepsilon}$ and $H_0^{\frac{1}{2}+
\varepsilon}$
are comparable, hence ${\cal D} (H_0^{\frac{1}{2}+\varepsilon})={\cal D}
(H_X^{\frac{1}{2}+\varepsilon}). \quad \Box$

The next theorem ensures the compatibility between the two charts in the
overlapping region ${\cal M}_0 \cap {\cal M}_X$.

\begin{theorem}
$\| \cdot \|_\varepsilon (X)$ and $\| \cdot \|_\varepsilon (0)$
are equivalent norms.
\end{theorem}
{\em Proof:} We need to show that there exist positive constants $m$
and $M$ such that $m\|Y\|_\varepsilon (0) \leq \|Y\|_\varepsilon
(X) \leq M\|Y\|_\varepsilon (0)$. We just write \begin{eqnarray*}
\|Y\|_\varepsilon (X) & = & \left\|R_X^{\frac{1}{2}+\varepsilon}H_0^
{\frac{1}{2}+\varepsilon}R_0^{\frac{1}{2}+\varepsilon}YR_0^{\frac{1}{2}
-\varepsilon}H_0^{\frac{1}{2}-\varepsilon}R_X^{\frac{1}{2}-\varepsilon}
\right\|\\
&\leq&\left\|R_X^{\frac{1}{2}+\varepsilon}H_0^{\frac{1}{2}+\varepsilon}
\right\|\left\|H_0^{\frac{1}{2}-\varepsilon}R_X^{\frac{1}{2}-\varepsilon}
\right\|\|Y\|_\varepsilon(0) \\
&=& M\|Y\|_\varepsilon(0)
\end{eqnarray*}
and, for the inequality in the other direction, we
write
\begin{eqnarray*}
\|Y\|_\varepsilon (0) & = & \left\|R_0^{\frac{1}{2}+\varepsilon}H_X^{
\frac{1}{2}+\varepsilon}R_X^{\frac{1}{2}+\varepsilon}YR_X^{\frac{1}{2}
-\varepsilon}H_X^{\frac{1}{2}-\varepsilon}R_0^{\frac{1}{2}-\varepsilon}
\right\| \\
&\leq&\left\|R_0^{\frac{1}{2}+\varepsilon}H_X^{\frac{1}{2}+\varepsilon}
\right\|\left\|H_X^{\frac{1}{2}-\varepsilon}R_0^{\frac{1}{2}-\varepsilon}
\right\|\|Y\|_\varepsilon(X) \\
  & = & m^{-1}\|Y\|_\varepsilon (X). \quad \Box
\end{eqnarray*}

We see that ${\cal T}_\varepsilon (0)$ and ${\cal T}_\varepsilon (X)$
are, in fact, the same Banach space furnished with two equivalent
norms, and observe that the quotient spaces ${\cal T}_\varepsilon
(0)/ \! \sim$ and ${\cal T}_\varepsilon (X)/ \! \sim$ are exactly the
same set. The general theory of Banach manifolds does the rest \cite{Lang}.

We continue in this way, adding a new patch around another point $\rho_{X'}$
in ${\cal M}_0$ or around some other point in ${\cal M}_X$ but
outside ${\cal M}_0$. Whichever point we start from, we get a
third piece ${\cal M}_X$ with chart into an open subset of the
Banach space $\left\{Y \in {\cal T}_\varepsilon (X'): \rho_{X'} \! \cdot \!
Y=0\right\}$, with norm $\|Y\|_\varepsilon (X'):=\left\|R_{X'}^{\frac{1}{2}+
\varepsilon}YR_{X'}^{\frac{1}{2}-\varepsilon}\right\|$
equivalent to the previously defined norms. We can go on
inductively, and all the norms of any overlapping regions will be
equivalent.

\begin{definition}
The information manifold ${\cal M}(H_0)$ defined by $H_0$ consists
of all states obtainable in a finite numbers of steps, by
extending ${\cal M}_0$ as explained above.
\end{definition}

These states are well defined in the following sense. If, for two
different sets of perturbations $X_1, \ldots , X_n$ and $Y_1, \ldots
,Y_m$, we have $X_1+ \cdots +X_n=Y_1 + \cdots +Y_m$ as forms on
${\cal D}(H_0^{\frac{1}{2}-\varepsilon})$, then we arrive at the
same state either taking the route $X_1, \ldots , X_n$
or taking the route $Y_1, \ldots ,Y_m$, since the self-adjoint
operator associated with the form $q_0+X_1+ \cdots +X_n=q_0+Y_1
+ \cdots +Y_m$ is unique.

\subsection{Affine Geometry in ${\cal M}(H_0)$}

The set $A=\left\{\widehat X \in \widehat{\cal T}_\varepsilon (0):
\widehat X=X-\rho_0 \! \cdot \! X, \|X\|_\varepsilon (0)<1-\beta_0\right\}$
is a convex subset of the Banach space $\widehat{\cal T}_\varepsilon (0)$
and so has an affine structure coming from its linear structure.
We provide ${\cal M}_0$ with an affine structure induced from $A$
using the patch $\widehat X \mapsto \rho_X$ and call this the
canonical or $(+1)$-affine structure. The $(+1)$-convex mixture of
$\rho_X$ and $\rho_Y$ in ${\cal M}_0$ is then $\rho_{\lambda X+(1-\lambda
)Y}$, $(0 \leq \lambda \leq 1)$, which differs from the previously defined
$(-1)$-convex mixture \mbox{$\lambda \rho_X +(1-\lambda )\rho_Y$}.

Given two points $\rho_X$ and $\rho_Y$ in ${\cal M}_0$ and their
tangent spaces $\widehat{\cal T}_\varepsilon (X)$ and $\widehat{\cal T}_
\varepsilon(Y)$, we define the $(+1)$-parallel transport $U_L$ of
$(Z-\rho_X \! \cdot \! Z) \in \widehat{\cal T}_\varepsilon (X)$
along any continuous path $L$ connecting $\rho_X$ and $\rho_Y$ in
the manifold to be the point $(Z-\rho_Y \! \cdot \! Z) \in \widehat
{\cal T}_\varepsilon(Y)$. Clearly $U_L(0)=0$ for every $L$, so the
$(+1)$-affine connection given by $U_L$ is torsion free. Moreover, $U_L$
is independent of $L$ by construction, thus the $(+1)$-affine connection
is flat. We see that the $(+1)$-parallel transport just moves the
representative point in the line $\{Z+\alpha I\}_{\alpha \in {\bf R}}$
from one hyperplane to another.

Now consider a second piece of the manifold, say ${\cal M}_X$. We
have the $(+1)$-affine structure on it again by transfer of
structure from $\widehat{\cal T}_\varepsilon (X)$. Since both $\widehat
{\cal T}_\varepsilon (0)$
and $\widehat{\cal T}_\varepsilon (X)$ inherit their affine
structures from the linear structure of the same set (either ${\cal T}
_\varepsilon (0)$ bor ${\cal T}_\varepsilon (X)$), we see that the
$(+1)$-affine structures of ${\cal M}_0$ and ${\cal M}_X$ are the same
on their overlap. We define the parallel transport in ${\cal M}_X$ again by
moving representative points around. To parallel transport a point
between any two tangent spaces in the union of the two pieces, we
proceed by stages. For instance, if $U$ denotes the parallel
transport from $\rho_0$ to $\rho_X$, it is straightforward to
check that $U$ takes a convex mixture in $\widehat{\cal T}_\varepsilon (0)$
to a convex mixture in $\widehat{\cal T}_\varepsilon (X)$. So, if $\rho_Y
\in{\cal M}_0$ and $\rho_{Y'} \in {\cal M}_X$ are points outside the
overlap, we parallel transport from $\rho_Y$ to $\rho_{Y'}$ following the
route $\rho_Y \to \rho_0 \to \rho_X \to \rho_{Y'}$.
Continuing in this way, we furnish the whole ${\cal M}(H_0)$ with a
$(+1)$-affine structure and a flat, torsion free, $(+1)$-affine connection.

Although each hood in ${\cal M}(H_0)$ is clearly $(+1)$-convex, we
have not been able to prove that ${\cal M}(H_0)$ is itself $(+1)$-convex.

\section{Analyticity of the Free Energy}

The free energy of the state $\rho_X=Z_X^{-1}e^{-H_X} \in {\cal C}_{\beta_X}
\subset {\cal M}, \beta_X
<1$, is the function $\Psi : {\cal M} \to {\bf C}$ given by
\begin{equation}
\Psi (\rho_X):=\log Z_X.
\end{equation}

We say that $Y$ is an $\varepsilon$-bounded direction if $Y \in {\cal T}_
\varepsilon(X)$. We now show that $\Psi (\rho_X)$ is infinitely Fr\'{e}chet
differentiable when the Fr\'{e}chet derivatives are taken in an
$\varepsilon$-bounded direction $\lambda V$ and that, in this case, it has
a convergent Taylor series for sufficiently small $\lambda$.

The $n$-th Fr\'{e}chet derivative of $\Psi_X \equiv \Psi (\rho_X)$ in
the $\varepsilon$-bounded directions $V_1, \ldots ,V_n$ is given
by $(n!)^{-1}$ times the Kubo $n$-point function \cite{Araki}
\begin{equation}
\mbox{Tr} \int_{0}^{1}d\alpha_1 \int_{0}^{1}d\alpha_2 \cdots \int_{0}
^{1}d\alpha_{n-1} [\rho_X^{\alpha_1}V_1\rho_X^{\alpha_2}V_2 \cdots
\rho_X^{\alpha_n}V_n],
\end{equation}
where $\alpha_n=1-\alpha_1 - \cdots - \alpha_{n-1}$.

We begin by estimating the trace of $[\rho_X^{\alpha_1}V_1\rho_X^
{\alpha_2}V_2 \cdots\rho_X^{\alpha_n}V_n]$ as written as
\begin{eqnarray*}
\lefteqn{[\rho_X^{\alpha_1 \beta_X}][H_X^{1-\delta_n + \delta_1}
\rho_X^{(1-\beta_X)\alpha_1}][R_X^{\delta_1}V_1R_X^{1-\delta_1}]
[\rho_X^{\alpha_2 \beta_X}][H_X^{1-\delta_1 + \delta_2}\rho_X^
{(1-\beta_X)\alpha_2}]} \\
 & & [R_X^{\delta_2}V_2R_X^{1-\delta_2}] \cdots [\rho_X^{\alpha_n
\beta_X}][H_X^{1-\delta_{n-1} +
\delta_n}\rho_X^{(1-\beta_X)\alpha_n}][R_X^{\delta_n}V_nR_X^{1-\delta_n}],
\end{eqnarray*}
with $\delta_j \in \left[\frac{1}{2}-\varepsilon ,\frac{1}{2}+\varepsilon
\right]$ to be specified soon. In this product, we have $n$ factors of
the form $[\rho_X^{\alpha_j \beta_X}]$, $n$ factors of the form $[R_X^
{\delta_j}V_jR_X^{1-\delta_j}]$, and $n$ factors of the form
$[H_X^{1-\delta_{j-1} + \delta_j}\rho_X^{(1-\beta_X)\alpha_j}]$,
with $\delta_0$ standing for $\delta_n$.

For the factors $[\rho_X^{\alpha_j \beta_X}]$, putting $p_j=1/\alpha_j$,
H\"{o}lder inequality leads to the trace norm
bound
\begin{equation}
\left\|[\rho_X^{\alpha_1 \beta_X}] \cdots [\rho_X^{\alpha_n
\beta_X}]\right\|_1 \leq
\left\|\rho_X^{\beta_X}\right\|_1^{\alpha_1} \cdots
\left\|\rho_X^{\beta_X}\right\|_1^{\alpha_n}=\left\|\rho_X^{\beta_X}\right\|
_1<\infty.
\label{estimate1}
\end{equation}

By virtue of lemma~\ref{monotonicity}, we know that the factors $[R_X^
{\delta_j}V_jR_X^{1-\delta_j}]$
are bounded in operator norm by

\begin{equation}
\left\|R_X^{\delta_j}V_jR_X^{1-\delta_j}\right\| \leq
\left\|R_X^{\frac{1}{2}+\varepsilon}V_jR_X^{\frac
{1}{2}-\varepsilon}\right\|=\left\|V_j\right\|_\varepsilon (X)<\infty.
\label{estimate2}
\end{equation}

In both these cases, the bounds are independent of
$\alpha$. The hardest case turns out to be the factors $[H_X^
{1-\delta_{j-1} +\delta_j}\rho_X^{(1-\beta_X)\alpha_j}]$, where the
estimate, as we will see, does depend on $\alpha$ and we have to worry
about integrability. For them, the spectral theorem gives the operator
norm bound

\begin{eqnarray}
\lefteqn{\left\|H_X^{1-\delta_{j-1} +
\delta_j}\rho_X^{(1-\beta_X)\alpha_j}\right\| =
Z_X^{-\alpha_j(1-\beta_X)} \sup_{x \geq 1}
\left\{x^{1-\delta_{j-1}+\delta_j}e^{-(1-\beta_X)\alpha_jx}\right\}}
\nonumber \\
&\leq & Z_X^{-\alpha_j(1-\beta_X)}\left(\frac{1-\delta_{j-1}+\delta_j}
{(1-\beta_X)\alpha_j}\right)^{1-\delta_{j-1}+\delta_j}e^{-(1-\delta_{j-1}+
\delta_j)}.
\label{bound}
\end{eqnarray}

Apart from $\alpha_j^{-(1-\delta_{j-1}+\delta_j)}$, the other
terms in (\ref{bound}) will be bounded independently of $\alpha$. To deal
with the integral of
$\alpha_j^{-(1-\delta_{j-1}+\delta_j)}d\alpha_j$, we divide the
region of integration in $n$ (overlapping) regions $S_j := \{\alpha:
\alpha_j \geq 1/n \}$
(since $\sum \alpha_j \! = \!1$). For the region $S_n$, for instance, the
integrability at $\alpha_j =0$ is guaranteed if we choose $\delta_j$
such that $\delta_j<\delta_{j-1}$. So we take $\delta_n=\delta_0 >
\delta_1 > \cdots >\delta_{n-1}$. We must have $\delta_j \in \left[\frac
{1}{2}-\varepsilon,\frac{1}{2}+\varepsilon\right]$, then we choose $\delta_n
=\frac{1}{2}+\varepsilon$,
$\delta_1 = \frac{1}{2}+\varepsilon - \frac{2\varepsilon}{n}$, $\delta_2 =
\frac{1}{2}+\varepsilon-\frac{4\varepsilon}{n}$, \ldots , $\delta_{n-1} =
\frac{1}{2}-\varepsilon +\frac{2\varepsilon}{n}$. Then each of the $(n-1)$
integrals, for $j=1, \ldots,n-1$,
is \[\int_{0}^{1}\alpha_j^{-(1-\delta_{j-1}+\delta_j)}d\alpha_j=
(\delta_{j-1}-\delta_j)^{-1}=\frac{n}{2\varepsilon}\]
resulting in a contribution of
$\left(\frac{n}{2\varepsilon}\right)^{n-1}$. The last integrand in
$S_n$ is \mbox{$\alpha_n^{-(1-\delta_{n-1}+\delta_n)} \leq n^2$}. The
same bound holds for the other regions $S_j ,j=1, \ldots ,n-1$,
giving a total bound \begin{equation}
\prod_{j=1}^{n} \int_{0}^{1}\alpha_j^{-(1-\delta_{j-1}+\delta_j)}
d\alpha_j \leq n\left[\frac{n^2n^{n-1}}{(2\varepsilon)^{n-1}}\right]=
\frac{n^2 n^n}{(2\varepsilon)^{n-1}}.
\label{estimate3}
\end{equation}

Now that we have fixed $\delta_j$, the promised bound for the other
terms in (\ref{bound}) is
\begin{eqnarray}
\lefteqn{\prod_{j=1}^{n}Z_X^{-\alpha_j(1-\beta_X)}\left(\frac{1-\delta
_{j-1}+\delta_j}{1-\beta_X}\right)^{1-\delta_{j-1}+\delta_j}} \nonumber \\
&\leq&4Z_X^{-(1-\beta_X)}(1-\beta_X)^{-n}e^{-n}
\label{estimate4}
\end{eqnarray}
since $(1-\delta_{j-1}+\delta_j)<1$ except for one term, when it
is less than 2.

Collecting the estimates
(\ref{estimate1}),(\ref{estimate2}),(\ref{estimate3}) and
(\ref{estimate4}), we get the
following bound for the $n$-point function \begin{equation}
4\left\|\rho_X^{\beta_X}\right\|_1 Z_X^{-(1-\beta_X)}
(2\varepsilon)n^2 n^n e^{-n}\prod_j\left[\frac
{\left\|V_j\right\|_\varepsilon \! (X)}{2\varepsilon(1-\beta_X)}\right].
\end{equation}

Thus $\Psi_X$ is infinitely Fr\'{e}chet differentiable in
$\varepsilon$-bounded direction and the Taylor series converges if
$\left\|V_j\right\|_\varepsilon \! (X)<(1-\beta_X)2\varepsilon$.
Notice that this is stronger than the condition that $\rho_{V+X}$
lies in a $\varepsilon$-hood of $\rho_X$.

Finally, let us say that a map $\Phi :{\cal U} \to {\bf C}$, on
a hood $\cal U$ in $\cal M$, is $(+1)$-analytic in $\cal U$ if
it is infinitely often Fr\'{e}chet differentiable in all
$\varepsilon$-bounded directions $\lambda V$ and $\Phi(\rho_X)$
has a convergent Taylor expansion for sufficiently small of
$\lambda$. In particular, the $(-1)$-coordinates $\eta_X=\rho \!
\cdot \! X$ (mixture coordinates) are analytic, since they are
derivatives of the free energy $\Psi_X$. This specification of the
sheaf of germs of analytic functions defines a real analytic
structure on the manifold.

\vspace{.1in}
\noindent{\bf Acknowledgments}

We thank E. B. Davies for help with lemma~\ref{domains}. RFS
thanks Rektor Jamiolkowski of Torun University for hospitality, May 1999, where this
work was started.


\begin{thebibliography}{99}

\bibitem{Amari} S.-I. Amari, {\bf Differential Geometric Methods in
Statistics}, {\em Lecture Notes in Statistics}, {\bf 28}, Springer-Verlag,
New York, 1985.

\bibitem{Araki} H. Araki, {\em Relative Hamiltonians for Faithful
Normal States of a von Neumann Algebra}, Publ. R. I. M. S.
(Kyoto), {\bf 9}, 165-209, 1968.

\bibitem{Davies} E. B. Davies, {\bf One-parameter semigroups},
Academic Press, 1980.

\bibitem{GibilescoPistone} P. Gibilesco and G. Pistone, {\em
Connections on nonparametric statistical manifolds by Orlicz space
geometry}, Inf. Dim. Analysis, Quant. Prob. and Rel. Top., {\bf
1}, 325-347, 1998.

\bibitem{Hasagawa1} H. Hasagawa, {\em $\alpha$-Divergence of the
Noncommutative Information Geometry}, Rep. Math. Phys., {\bf 33},
87-93, 1993.

\bibitem{Hasagawa2} H. Hasagawa, {\em Noncommutative Extension of
the Information Geometry}, in {\bf Quantum Communication and
Measurement}, eds. V. P. Balavkin, O. Hirota and R. L. Hudson,
Plenum Press, 1995.

\bibitem{Kass} R. A. Kass and P. W. Vos, {\bf Geometric
Foundations of Asymptotic Inference}, Wiley, New York, 1997.

\bibitem{Kato} T. Kato, {\bf Perturbation Theory for Linear
Operators}, Springer-Verlag, 1966.

\bibitem{Lang} S. Lang, {\bf Differential and Riemannian
Manifolds}, Springer-Verlag, 1995.

\bibitem{Murray} M. K. Murray and J. W. Rice, {\bf Differential
Geometry and Statistics}, {\em Monographs on Statistics and Applied
Probability}, {\bf 48}, Chapman \& Hall, 1993.

\bibitem{Nagaoka} H. Nagaoka, {\em Differential Geometric Aspects
of Quantum State Estimation and Relative Entropy}, in {\bf Quantum Communication and
Measurement}, eds. V. P. Balavkin, O. Hirota and R. L. Hudson,
Plenum Press, 1995.

\bibitem{Petz} D. Petz, {\em Geometry of Canonical Correlation on
the State Space of a Quantum System}, J. Math. Phys., {\bf 35},
780-795, 1994.

\bibitem{PetzSudar} D. Petz and C.Sudar, {\em Geometries of Quantum
States}, J. Math. Phys., {\bf 37}, 2662-2673, 1996.

\bibitem{PetzToth} D. Petz and G. Toth, {\em The Bogoliubov inner
product in quantum statistics}, Lett. Math. Phys., {\bf 27},
205-216, 1993.

\bibitem{Pietsch} A. Pietsch, {\bf Nuclear Locally Convex Spaces},
Springer-Verlag, 1972.

\bibitem{PistoneSempi} G. Pistone and C. Sempi, {\em An
infinite dimensional geometric structure on the space of all
probability measures equivalent to a given one}, Ann. Stat., {\bf
33}, 1543-1561, 1995.

\bibitem{ReedSimon} M. Reed and B. Simon, {\bf Methods of Modern
Mathematical Physics}, vol. 2, Academic Press, 1975.

\bibitem{Streater1} R. F. Streater, {\em The Information Manifold
for Relatively Bounded Potentials}, to appear in the Bogoliubov Memorial Volume, ed. A. A. Slavnov, Stecklov Institute.

\bibitem{Streater2} R. F. Streater, {\em The Analytic Quantum
Information Manifold}, to appear in {\bf Stochastic Processes, Physics and Geometry}, eds. F. Gesztesy,
S. Paycha and H. Holden, Canad. Math. Soc.

\end{thebibliography}
\end{document}